\documentclass[aip, amsmath, amssymb, reprint]{revtex4-2}

\usepackage{graphicx}
\usepackage{dcolumn}
\usepackage{hyperref}
\usepackage{bm}
\usepackage[utf8]{inputenc}
\usepackage[T1]{fontenc}
\usepackage{mathptmx}
\usepackage{color}

\begin{document}

\preprint{AIP/123-QED}

\title[Bayesian Phase Estimation via Active Learning]
{Bayesian Phase Estimation via Active Learning}

\author{Yuxiang Qiu}
\affiliation{Guangdong Provincial Key Laboratory of Quantum Metrology and Sensing $\&$ School of Physics and Astronomy, Sun Yat-Sen University (Zhuhai Campus), Zhuhai 519082, China}
\affiliation{State Key Laboratory of Optoelectronic Materials and Technologies, Sun Yat-Sen University (Guangzhou Campus), Guangzhou 510275, China}

\author{Min Zhuang}
\affiliation{Guangdong Provincial Key Laboratory of Quantum Metrology and Sensing $\&$ School of Physics and Astronomy, Sun Yat-Sen University (Zhuhai Campus), Zhuhai 519082, China}
\affiliation{State Key Laboratory of Optoelectronic Materials and Technologies, Sun Yat-Sen University (Guangzhou Campus), Guangzhou 510275, China}

\author{Jiahao Huang}
\altaffiliation{Email: hjiahao@mail2.sysu.edu.cn, eqjiahao@gmail.com}
\affiliation{Guangdong Provincial Key Laboratory of Quantum Metrology and Sensing $\&$ School of Physics and Astronomy, Sun Yat-Sen University (Zhuhai Campus), Zhuhai 519082, China}
\affiliation{State Key Laboratory of Optoelectronic Materials and Technologies, Sun Yat-Sen University (Guangzhou Campus), Guangzhou 510275, China}

\author{Chaohong Lee}
\altaffiliation{Email: lichaoh2@mail.sysu.edu.cn, chleecn@gmail.com}
\affiliation{Guangdong Provincial Key Laboratory of Quantum Metrology and Sensing $\&$ School of Physics and Astronomy, Sun Yat-Sen University (Zhuhai Campus), Zhuhai 519082, China}
\affiliation{State Key Laboratory of Optoelectronic Materials and Technologies, Sun Yat-Sen University (Guangzhou Campus), Guangzhou 510275, China}

\date{\today}

\begin{abstract}
  Bayesian estimation approaches, which are capable of combining the information of experimental data from different likelihood functions to achieve high precisions, have been widely used in phase estimation via introducing a controllable auxiliary phase. 
  Here, we present a non-adaptive Bayesian phase estimation (BPE) algorithms with an ingenious update rule of the auxiliary phase designed via active learning. 
  Unlike adaptive BPE algorithms, the auxiliary phase in our algorithm is determined by a pre-established update rule with simple statistical analysis of a small batch of data, instead of complex calculations in every update trails.
  As the number of measurements for a same amount of Bayesian updates is significantly reduced via active learning, our algorithm can work as efficient as adaptive ones and shares the advantages (such as wide dynamic range and perfect noise robustness) of non-adaptive ones.
  Our algorithm is of promising applications in various practical quantum sensors such as atomic clocks and quantum magnetometers. 
\end{abstract}

\maketitle

\section{\label{sec:1 introduction}introduction}

Quantum phase estimation is at the core of precision measurement and sensing~\cite{Giovannetti2006, Giovannetti2011, Gross2012,Degen2017,Braun2018, Pezze2018}.
The estimation of an unknown phase via interferometric techniques are widely used in quantum sensors~\cite{Lane1993,Tsang2012,Giovannetti2012,Waldherr2012,Pezze2015} such as atomic clocks, magnetometers, and gravimeters.
Generally, there are two different approaches to accomplish the phase estimation problem: frequentist and Bayesian.
Frequentist and Bayesian phase estimation strategies lead to conceptually different information on the estimated parameters and their uncertainties according to the results of measurements~\cite{Li2018}.
Compared to conventional frequentist estimation approaches~\cite{Kay1993,Lehmann1998}, the Bayesian approach is capable to obtain information from every single measurement output.
The Bayesian approach makes use of the Bayes' theorem to update the posterior probability, which describes the current knowledge about the random variable based on the available measurement results. 
This allows the Bayesian approach to provide statistical information for any number of measurements.

Bayesian phase estimation (BPE) is known to be particularly efficient and versatile. 
%They are particularly well-suited for quantum information, owing to their generality, robustness and ease of incorporating likelihood information into the prior confidence. 
%
In recent, BPE protocol becomes a good choice on account of its ability to reduce the measurement repeats needed while preserving the robustness against noises~\cite{Paesani2017, Wiebe2016, Wang2017, Ruster2017}. 
It is of great value in practical application where only a limited number of measurements are available~\cite{Rubio2018,Rubio2019}.
Generally, BPE algorithms can be classified into adaptive (online)~\cite{Wiseman1995,Berry2002,Armen2002,Paesani2017,Lumino2018,Dimario2020} and non-adaptive (offline) algorithms~\cite{Higgins2009,Berry2000,Said2011,Nusran2012,Nusran2014}.
In adaptive BPE algorithms, the auxiliary phases (or other controlled quantities) are calculated during the process of experiments by taking into account the previous measurement data. 
While in non-adaptive BPE algorithms, the auxiliary phases for each Bayesian update step is pre-determined in advance.
Obviously, adaptive algorithms inevitably require laborious calculations and operations in order to find step-wise optimized auxiliary phases. 
Differently, the auxiliary phases for non-adaptive ones are pre-determined~\cite{Said2011}.
Besides, non-adaptive algorithms generally require lower measurement fidelity, and show better dynamic range and greater consistency in sensitivity~\cite{Nusran2014}.
However, non-adaptive algorithms may require more measurement times so that the total experimental duration may be much longer. 
To benefit from non-adaptive BPE algorithms, can we effectively reduce the required measurement times?

Machine learning, which involves various algorithms and modeling tools for data processing, has been widely used in the fields of quantum science and technologies~\cite{RevModPhys.91.045002}.  
It provides a powerful tool for understanding and exploiting quantum effects, such as classifying many-body quantum phases~\cite{Carrasquilla2017,VanNieuwenburg2017}, speeding up many-body quantum simulations~\cite{PhysRevB.97.205140, PhysRevB.95.041101} and improving the performances of quantum sensors~\cite{Wang2017, Lumino2018, Schuff_2020}. 
In particular, active learning, which involves human intervention in data preprocessing, is a promising technique to solve the time- or resources-demanding problems~\cite{PhysRevLett.124.140504, PhysRevResearch.2.013287}.
The main idea in active learning is that, if a learning algorithm may choose the data that worth to learn from, it can perform better than traditional methods with substantially less data.
Here, in order to enjoy the advantages of non-adaptive algorithms and meanwhile reduce the required measurement times, we propose a BPE algorithm via active learning, which can select significant data in a pre-learning process to increase the efficiency.  
Our algorithm is not only as maneuverable as the non-adaptive algorithms, but also can provide a desirable measurement precision with a reduced number of measurement times.

In this article, we show how to combine non-adaptive BPE algorithm with active learning to estimate a phase with reduced measurement times.
Compared with conventional BPE algorithms, our algorithm can save up to $85\%$ measurement times.
Our numerical simulations show that the performances of the error and the uncertainty versus the Bayesian update times keep the same level achieved by conventional BPE algorithms. 
The uncertainty may reach the Ghosh bound of BPE, which scales as the standard quantum limit. 
The reduction of the required measurement times and the pre-determination of the variation of auxiliary phases makes our algorithm as efficient as the adaptive ones~\cite{Lumino2018}.
Moreover, our algorithm shows good dynamic range and robustness against noises.

\begin{figure*}[t]
  \includegraphics[width = 2.06\columnwidth]{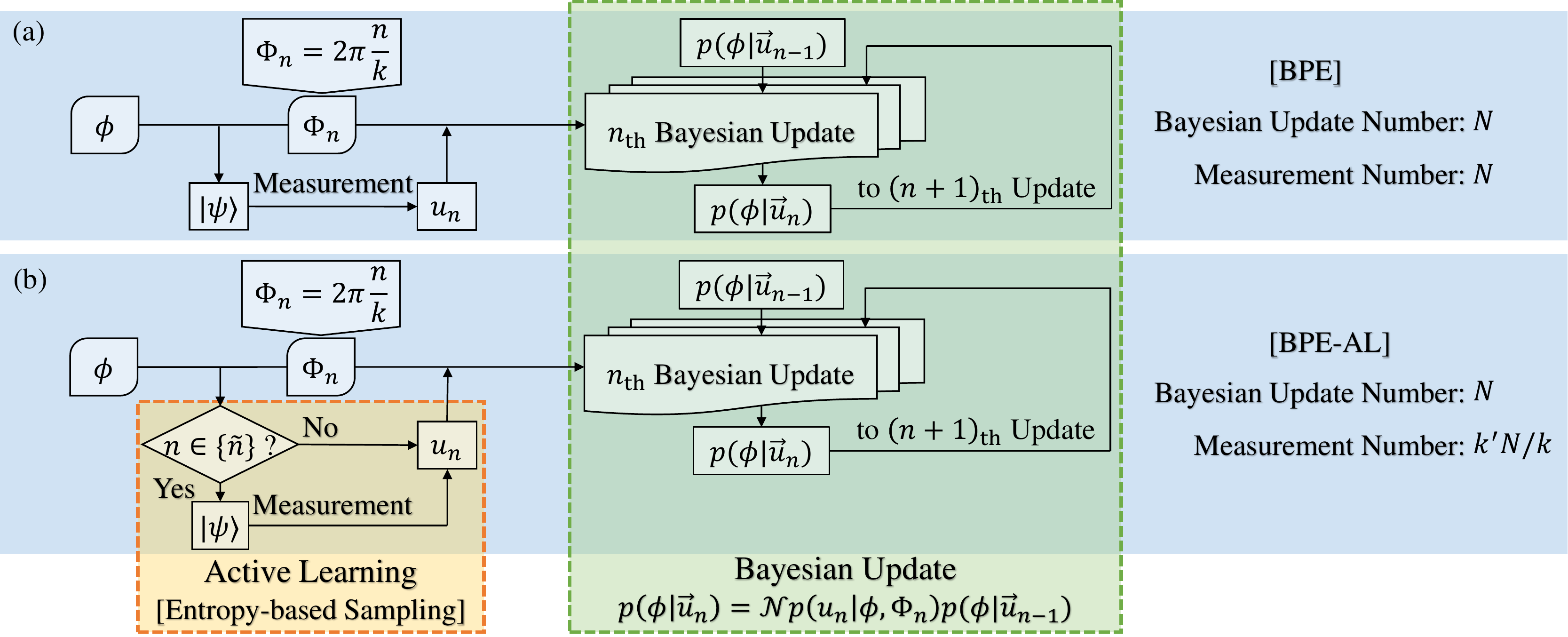}
  \caption{\label{fig_1_Bayes}
  The schematic diagrams for (a) the conventional BPE algorithm and (b) our BPE algorithm via active learning (labeled as BPE-AL).
  (a) In the $n$-th update, the auxiliary phase $\Phi_n$ is chosen as $2\pi n/k$, and then a measurement is implemented. The measurement outcome data $u_n$ is fed for Bayesian update to transform the prior probability into the posterior probability. Then, the posterior is used as the prior for the next Bayesian update. For the conventional BPE, the measurement times equals the Bayesian update times $N$. 
  (b) By using the active learning based on entropy-based sampling, our BPE-AL only need to select the significant auxiliary phases $\Phi_n$ with $n\in\{\tilde{n}\}$ for measurement. Thus, the measurement times can be reduced to $k'N/k$, where $k'$ can be much smaller than $k$.
  }
\end{figure*}

\begin{figure*}[ht]
  \includegraphics[width = 2.06\columnwidth]{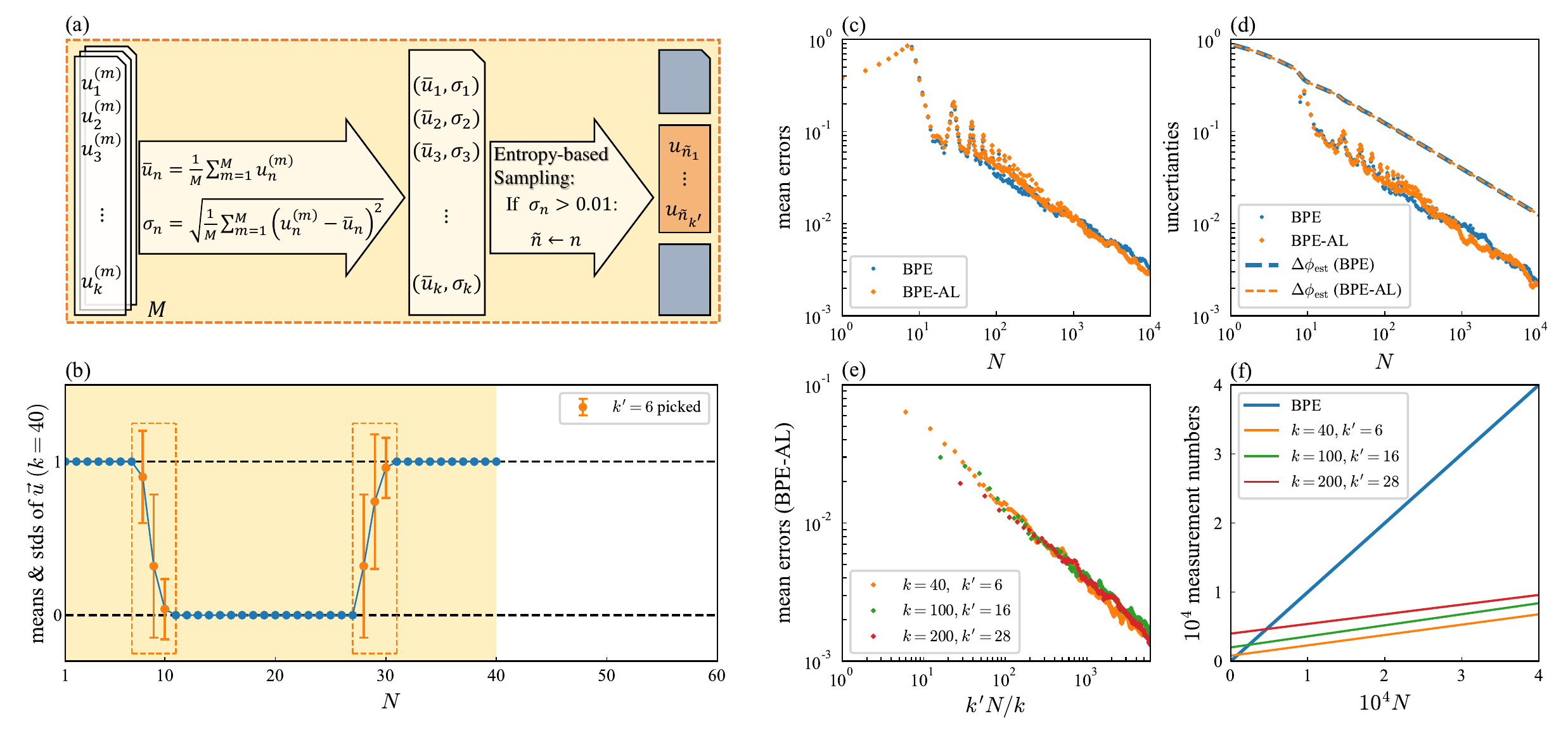}
  \caption{\label{fig:2}
  (a) The sketch of active learning process in our BPE-AL algorithm. From $M$ independent repeats of $k$ pre-estimation, the means $\bar{u}_n$ and its stds $\sigma_n$ of the resultant data are calculated. By using the entropy-based sampling, $k'$ data with significant stds are picked out, as indexed by $\tilde{n}_1$ to $\tilde{n}_{k'}$ in the orange box.
  (b) Results of the active learning process with $k=40$ and $M=20$. Blue points and errorbars correspond to $\bar{u}_n$ and $\sigma_n$, respectively. The data with significant stds are highlighted by orange color, see the orange dashed rectangles.
  (c) The mean errors obtained from $50$ independent simulations. Blue points are results of the conventional BPE algorithm, while orange diamonds stand for the results of our BPE-AL algorithm with $k=40$ and $k'=6$. 
  (d) The associated stds and uncertainties are plotted by points and dashed lines respectively. The uncertainties are calculated by the stds of the $N$-th posterior probability.
  (e) Scalings of mean errors with our BPE-AL for $k=40, 100, 200$ (respectively represented by orange, green and red diamonds) versus the measurement times $N_{\mathrm{meas}} \approx k'N/k$.
  (f) The measurement times versus the Bayesian update times $N$. Blue line corresponds the conventional BPE, while orange, green and red lines respectively correspond to our BPE-AL with $k=40$, 100 and 200.
  }
\end{figure*}

\section{\label{sec:2 Algorithm}Algorithm}

\subsection{\label{sec:2-1 BPE}General Procedure of Bayesian Phase Estimation}

In a general quantum phase estimation procedure, a probe state $|\psi\rangle$ undergoes a transformation to an output state $|\psi(\phi)\rangle$ that depends on an unknown phase $\phi$. 
The goal is to estimate $\phi$ according to the measurement results on $|\psi(\phi)\rangle$. 
To perform the Bayesian phase estimation, one may introduce an auxiliary phase $\Phi$ to adjust the probability distribution. 
Thus, a measurement outcome $u$ occurs with probability $p(u|\phi,\Phi)=\langle\psi(\phi,\Phi)| \hat{\Lambda}_{u} |\psi(\phi,\Phi)\rangle$, where $\psi(\phi,\Phi)$ is the output state and $\hat{\Lambda}_u$ is a POVM operator~\cite{wiseman2009quantum}. 
In most situations, the probability $p(u|\phi,\Phi)$ is a periodic function of the unknown phase $\phi$.

The simplest and most widely used example of quantum phase estimation is the Ramsey or Mach-Zehnder interferometry with individual two-mode particles~\cite{Ramsey1986,Lee2012,Settles2012,Bonato2016,Danilin2018}.
The output state can be expressed as $|\psi(\phi, \Phi)\rangle=(e^{-i\Phi}|0\rangle+e^{-i\phi}|1\rangle)/\sqrt{2}$~\cite{Wiebe2016,Dinani2019,Holland1933,Lee2006,Cronin2009,Lumino2018,Zheng2020,Ou2020}, where $|0\rangle$ and $|1\rangle$ respectively represent the two modes. The auxiliary phase $\Phi$ can be well controlled as desired~\cite{Hradil1996,Said2011,Paesani2017,Lumino2018,Rambhatla2020}. 
After the operation of recombination, one can finally get the probability of finding the state in $|u\rangle$ (whose measurement outcome $u$ equals 0 or 1) as~\cite{Higgins2009,Wiebe2016,Lumino2018,Dinani2019}:
\begin{equation}\label{Likelihood}
	p(u|\phi,\Phi)=\frac{1}{2}\left[1+(-1)^{u}\cos{(\phi-\Phi)}\right].
\end{equation}

In single-shot measurements, a binary outcome $u\in\{0,1\}$ can always be obtained. 
For an ensemble of $R$ particles ($R>1$), the binary measurement outcome can be obtained via the threshold method~\cite{Dinani2019,Nusran2012,DAnjou2014}. 
The threshold measurement only needs to tell the particle occupation on which state is quantitatively surpass the particles on the other state, instead of preparing a system available for single shot measurements.
If $r$ particles are found in $|1\rangle$, then the resultant outcome $u$ is given by
\begin{equation}\label{threshold_method}
  u=\left\lbrace \begin{array}{l}
    0,\;\quad\;r\le R/2, \\
    1,\;\quad\;r>R/2.
  \end{array}\right.
\end{equation}
This refers to the ``majority voting'', which is widely achieved by analysis of spin-dependent photo-luminescence data intensity~\cite{Dinani2019,Gupta2016} or voltage signals, or equivalently by repetitive measurements upon a single-spin system, which is also usually implemented in real experiments.

The BPE algorithms involve a set of measurements and updating the prior probability distribution according to the Bayes' theorem~\cite{linden_dose_toussaint_2014, Spagnolo2019}.
Given the first $n-1$ outcomes $\vec{u}_{n-1}=(u_1,u_2,\ldots,u_{n-1})$, the posterior probability distribution
\begin{equation}\label{Bayesian_update}
  	p(\phi|{\vec{u}}_n)=\mathcal{N}p(u_n|\phi,\Phi_n)p(\phi|\vec{u}_{n-1}),
\end{equation}
where $p(\phi|\vec{u}_{n-1})$ is the prior probability, $p(u_n|\phi,\Phi_n)$ is the likelihood and $\mathcal{N}=[\int{p(u|\phi,\Phi)p(\phi)}]^{-1}$ is a normalization factor.

The auxiliary phase for the $n$-th iteration $\Phi_{n}$ can be designed by non-adaptive algorithms.
For the likelihood function given by Eq.(\ref{Likelihood}), the variation of the auxiliary phase can be designed with equal steps~\cite{Berry2000,Said2011,Nusran2014}, i.e., 
\begin{equation}\label{Phi_updating_rule}
  	\Phi_n=n\cdot\frac{2\pi}{k},
\end{equation}
where $k$ denotes the number of auxiliary phases in a period. 
This updating rule of $\Phi_n$, which uses pre-established measurement settings or policies~\cite{Hentschel2010,Hentschel2011,Rambhatla2020}, avoids expensive calculations and over-fast experimental rate~\cite{Wiebe2016}.  
The value of $k$ can be properly chosen within the permission of the experimental adjusting precision. 

As shown in FIG.\ref{fig_1_Bayes}(a), the phase estimation procedure for each Bayesian update can be described as follows.
\begin{itemize}
  \item Step 1: Given a prior probability distribution $p(\phi|\vec{u}_{n})$ [The initial $p(\phi|u_0)$ can be set as a uniform distribution over interval $[0, 2\pi)$].
  \item Step 2: Perform the experiment with auxiliary phase $\Phi_n$ according to Eq.(\ref{Phi_updating_rule}), and record the measurement outcome $u_n$.
  \item Step 3: Update the posterior probability distribution according to Eq.(\ref{Bayesian_update}).
  \item Step 4: Evaluate expectation $\phi_{\mathrm{est}}=\int \phi p(\phi|\vec{u}_n) d\phi$ and associated uncertainty $\Delta\phi_{\mathrm{est}}=\sqrt{\int \phi^2 p(\phi|\vec{u}_n) d\phi - \phi_{\mathrm{est}}^2}$ via the posterior probability $p(\phi|\vec{u}_n)$~\cite{Ruster2017,Lumino2018,Rubio2018,Rubio2019}.
\end{itemize}
For the next update, the current posterior probability distribution $p(\phi|\vec{u}_{n})$ is regarded as a new prior probability. Then return to Step 1 to start a new cycle until $n=N$.

This iteration makes BPE approaches different from traditional frequentist approaches~\cite{Li2018}, and allows better efficiency and noise robustness~\cite{Paesani2017}.
The key issue is the updating rule of auxiliary phase $\Phi$ because the algorithm cannot get any new useful information when the value of $\Phi$ is fixed. 
Adaptive algorithms make it efficiently feasible by designing particular updating rules or policies of $\Phi$~\cite{Berry2002,Hentschel2010,Nusran2014,Wiebe2016,Lumino2018,Rambhatla2020}.
While our algorithm is based upon the above non-adaptive procedure, as shown in the following, it utilizes the ideas of active learning to reduce the actual measurement times.

\subsection{\label{sec:2-2 BPE-AL}Bayesian Phase Estimation via Active Learning}

As a concept from machine learning, active learning~\cite{Settles2012,Carleo2019} involves a learning algorithm that can choose the data it worth to learn from. 
It can perform better than traditional methods with substantially less data.
Here we adopt the so-called \emph{entropy-based sampling}~\cite{Li2020} from active learning to select the measurement data that significantly affect the results. 
The entropy of a discrete probability distribution is defined as~\cite{Shannon2001} $H= -\sum_{i=1}^n\mathrm{P}(x_i)\log{\mathrm{P}(x_i)}$, where ${x_i}$ refers to all possible values of a random variable $x$, $P(x_i)$ is the probability it occurs.
The data with larger entropy generally contain more information~\cite{Gray2011,Ruster2017}.
Here, in our algorithm, we introduction a learning process to select the specific data with large entropy. 

In the learning process, we repeat the measurements $M$ times, where each repeat contains $k$ times of the Bayesian update~(\ref{Bayesian_update}) with $\Phi_n = 2\pi n/k$ and $n=\{1,2,\cdots, k\}$.
The measurement times $Mk$ in the learning process is usually much smaller than the Bayesian update times $N$ in the whole experiment. 
The measurement data $\{u_1^{(m)}, u_2^{(m)}, \ldots , u_k^{(m)}\}_{m=1,2,\ldots,M}$ are collected and shown in FIG.~\ref{fig_1_Bayes}~(b).
For the data obtained from the $n$-th update using $\Phi_n$, the associated entropy is given by $H_n=-\sum_{u=0,1}\mathrm{P}_n(u)\log{\mathrm{P}_n(u)}$. 
From this definition, one can find that when $P_n(0)=P_n(1)=1/2$, the entropy reaches its maximum. 
While when $\{P_n(0)~=~0,~P_n(1)~=~1\}$ or $\{P_n(0)~=~1,~P_n(1)~=~0\}$, the entropy equals zero. 
For the binary measurement outcomes, the standard deviation (std) has similar property of the entropy. 
Therefore selecting the phase $\Phi_n$ bringing significant std is equivalent to picking out the data with large entropy. 
The std can be calculated by the primal definition: 
\begin{equation}\label{std}
  	\sigma_n = \sqrt{\frac{1}{M}\sum_{m=1}^M (u_n^{(m)}-\bar u_n)^2},
\end{equation}
where the arithmetic mean $\bar u_n = \frac{1}{M} \sum_{m=1}^M u_n^{(m)}$.
Our algorithm aims to search for those $\Phi_n$ associated with $\sigma_n>0.01$, as shown in FIG.~\ref{fig:2}~(a).

The above active learning process will make the BPE procedure more efficient. 
As shown in FIG.~\ref{fig:2}~(b), through the active learning, the informative data (the orange points) with $\sigma_n>0.01$ (which are labelled as $\{\tilde{n}_1, \tilde{n}_2, ..., \tilde{n}_{k'}\}$) are selected. 
The other data with small std (the blue points) are regarded as uninformative.
Owing to the periodicity, we can deduce that the measurement data obtained after a Bayesian update using $\Phi_{\tilde{n}}$ are more informative than the others. 
Therefore the indices can be explicitly written as 
\begin{equation}\label{indices}
  	\{\tilde{n}\} = 
    \{ck+\tilde{n}_1, ck+\tilde{n}_2, \ldots, ck+\tilde{n}_{k'}\}_{c=1,2,...}.
\end{equation}
As shown in the light orange box in FIG.~\ref{fig_1_Bayes}~(b), we only perform real measurements for these $\Phi_{\tilde{n}}$ and record the measurement data $u_{\tilde{n}}$.
While for other $\Phi_{n}$ with $n \notin \{\tilde{n}\}$, we do not perform real measurements and the corresponding outcomes are given as $0$ or $1$ according to the results of learning.

By including the entropy-based sampling operation, the procedure of our BPE-AL algorithm is as following.
\begin{itemize}
  \item Step 0: Acquire the informative set $\{\tilde{n}\}$ via the entropy-based sampling method.  
  \item Step 1: Given a prior probability distribution $p(\phi|\vec{u}_{n})$ [The initial $p(\phi|u_0)$ can be set as a uniform distribution over interval $[0, 2\pi)$].
  \item Step 2: Obtain the data $u_n$ via a real measurement with the auxiliary phase $\Phi_n$ only if $n \in \{\tilde{n}\}$. Otherwise the measurement is canceled and the data is given by the results of learning.
  \item Step 3: Update the posterior probability distribution according to Eq.(\ref{Bayesian_update}).
  \item Step 4: Evaluate the expectation $\phi_{\mathrm{est}}$ and the associated uncertainty $\Delta\phi_{\mathrm{est}}$.
\end{itemize}
The difference between BPE and BPE-AL only occurs in the Step 2, where we add a conditional statement to decide whether a real measurement needs to be implemented.
Through picking out the most informative data with limited measurement times, the BPE-AL algorithm is capable to perform the procedure more fast. 
In this way, for the same Bayesian update times, the required measurement times can be observably smaller than the one for the conventional BPE. 
Thus, our algorithm equips the advantages of non-adaptive algorithms and meanwhile dramatically reduces the real measurement times.
In the following, we will show the performance analysis of our BPE-AL.

\section{\label{sec:3}Performance analysis}

\subsection{\label{sec:3-1}Measurement Times and Measurement Precision}

We discuss the measurement times at first. 
For the learning process, we set $M=20$, $k=40$, and the estimated phase $\phi=2.7624$ for demonstration. 
As shown in FIG.\ref{fig:2}~(b), the mean $\bar u_n$ and its std $\sigma_n$ are marked by points and errorbars, respectively. 
Particularly, the informative data whose $\sigma_n>0.01$ are highlighted by orange color.  
We find that, in a period of $\Phi_n$, only $k'=6$ informative data are selected from $k=40$ Bayesian updates. 
Thus, after the learning process, for every period (which includes $k=40$ Bayesian updates), only $k'=6$ real measurements are needed to performed.  
For the conventional BPE, the measurement times equals the Bayesian update times, $N_{\mathrm{meas}}=N$. 
By comparison, our BPE-AL algorithm can reduce the measurement times to $N_{\mathrm{meas}}=Mk+k'(N-Mk)/k$.
Roughly, if the Bayesian update times is sufficiently large $N \gg Mk$, the measurement times can be reduced to $N_{\mathrm{meas}} \approx k'N/k$.
Thus, the measurement times performed in experiments can be greatly reduced. 
For example, for $k=40$, $k'=6$, the measurement times of BPE-AL is reduced to $15\%$. 
Despite the measurement times decreases substantially, the Bayesian update times remains the same, which guarantee the measurement precision.

The comparison of measurement precisions with BPE and BPE-AL are shown in FIG.\ref{fig:2}~(c) and (d). 
We evaluate the absolute error $|\phi_{\mathrm{est}}-\phi|$ and the uncertainty $\Delta\phi_{\mathrm{est}}$.
To avoid possible numerical errors, the phase $\phi$ is discretized to $10^6$ points within a period $[0, 2\pi)$.
Based upon the results of $50$ independent simulation, we compare the mean errors $\frac{1}{50} \sum_{k=1}^{50} |\phi_{\mathrm{est}}^{(k)}-\phi|$ and the corresponding stds $\sqrt{ \frac{1}{50} \sum_{k=1}^{50} \left[ |\phi_{\mathrm{est}}^{(k)}-\phi| - \left( \frac{1}{50} \sum_{k=1}^{50} |\phi_{\mathrm{est}}^{(k)}-\phi| \right) \right]^2 }$ for the conventional BPE and our BPE-AL. 
The mean errors and uncertainties of our BPE-AL are all as good as the ones of the conventional BPE.
This indicates that our BPE-AL is valid for phase estimation with high-precision as the BPE does.

Further, we give the simulation results via our BPE-AL with different $k$, as shown in FIG.\ref{fig:2}~(e). 
There are almost no difference between the performances for different values of $k$ because the decreasing ratios of measurement times $k'/k$ are similar.
Under the condition of similar measurement precision, the measurement times required by our BPE-AL is much smaller compared with the one of the conventional BPE.
The measurement times versus the Bayesian update times $N$ via BPE and BPE-AL are shown in FIG.\ref{fig:2}~(f).
The reduction of measurement times can save much experimental time.

\subsection{\label{sec:3-2}Dynamic Range} 
\begin{figure}[t]
  \includegraphics[width = \columnwidth]{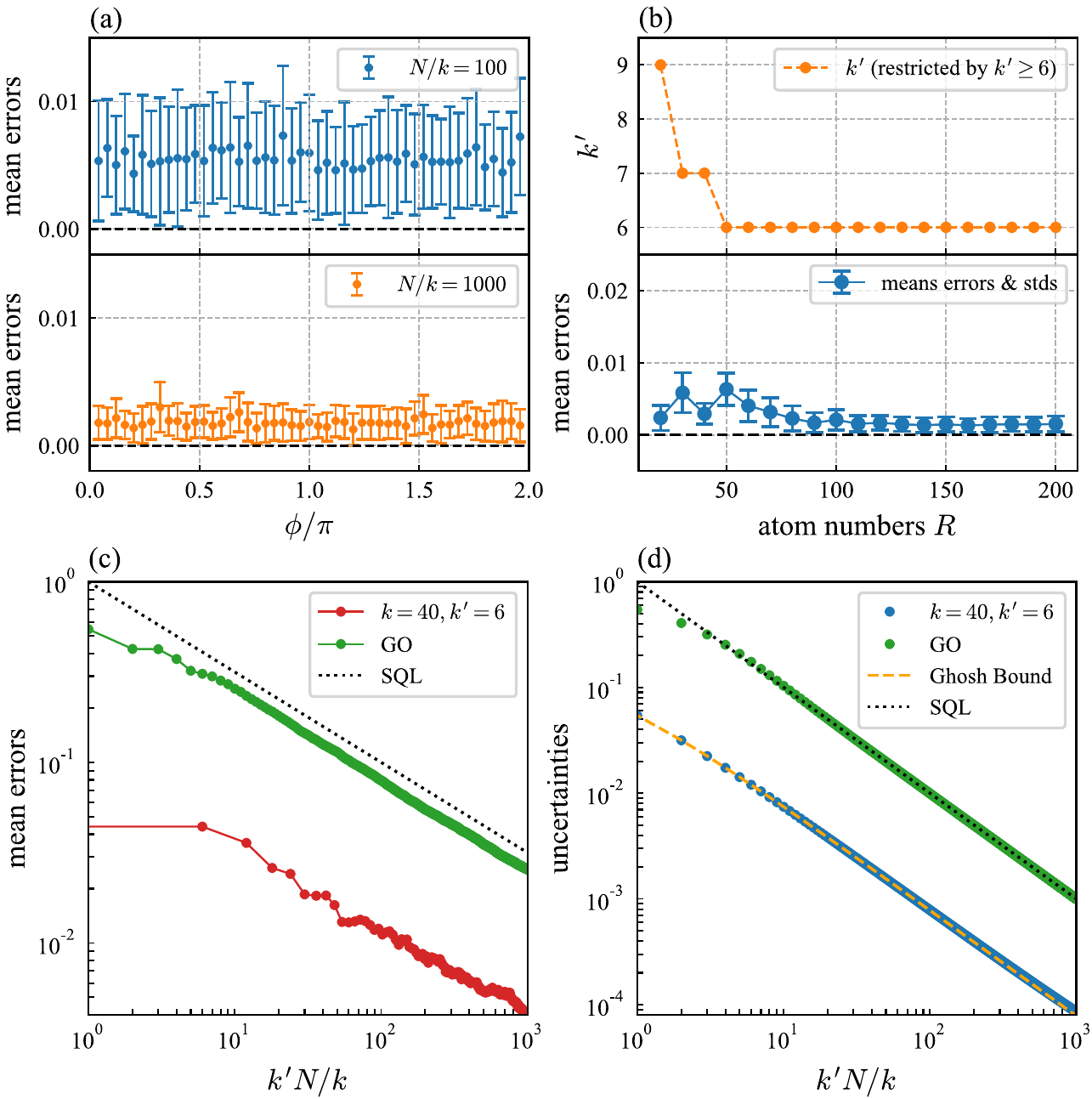}
  \caption{\label{fig:FIG3}
  (a) Mean errors and stds of $50$ independent simulations versus the estimated phases $\phi$ within $(0, 2\pi)$. Blue and orange points with errorbars correspond to our BPE-AL algorithm with $N/k=100$ and $N/k=1000$, respectively.
  (b) Mean errors and stds for $\phi=2.7624$ versus the particle number $R$ (from $20$ to $200$). The orange dotted dashed line stands for the measurement times $k^{\prime}$ in each $k=40$ iterations. 
  (c) The squared uncertainties $\Delta^2\phi_{\mathrm{est}}$ with our BPE-AL algorithm (blue points) versus the measurement times $k^{\prime}N/k$, while green points are squared uncertainties obtained by GO algorithm. The Ghosh bound is drawn with orange dashed line, and the standard quantum limit (SQL) is plotted as a black dashed line.
  (d) Mean errors of GO algorithm and our BPE-AL algorithm are shown as green and red dotted lines respectively. Results with our BPE-AL algorithm are reported in every $k'=6$ points to keep the same consumption of particle number.}
\end{figure}

The dynamic range~\cite{Waldherr2012, Nusran2012} of a physical quantity is its maximum value that can be detected with high precision. 
The ambiguity-free dynamic range of a phase is fundamentally limited to $2\pi$.
%Dynamic range~\cite{Waldherr2012, Paesani2017} describes the difference in estimations of BPE algorithm in different sub-intervals of $[0,2\pi)$. 
%
Usually, for adaptive BPE algorithms, the cosine likelihood function (\ref{Likelihood}) can cause inadequately convolution of the probability distribution functions that may result in bad performance~\cite{Paesani2017} for $\phi$ near $0$ or $2\pi$.
Luckily, our BPE-AL algorithm preserves the  dynamic range of non-adaptive BPE algorithms~\cite{Nusran2014}, which works well for all phases in the interval $[0, 2\pi)$.
We perform our BPE-AL algorithm to estimate different unknown phases $\phi$ in the interval $[0, 2\pi)$ and obtain their mean errors and uncertainties from $50$ independent simulations, see FIG.~\ref{fig:FIG3}(a).
For all phases in $[0,2\pi)$, their mean errors and uncertainties are similar without significant changes. 
This indicates that our BPE-AL algorithm can provide effective estimation for a unknown phase $\phi$ over the whole range of $[0,2\pi)$.

\subsection{\label{sec:3-3}Particle Number}

The particle number $R$ may influence the binary outcomes through the threshold criteria~(\ref{threshold_method}) and so that it will have an influence on the final results.  
Generally, a larger particle number $R$ will result in more precise binary outcomes. 
The results from different particle numbers $R$ are shown in FIG.~\ref{fig:FIG3}(b) for $N/k=1000$.
It is shown that our BPE-AL algorithm is robust against a lack of particles in each iteration, whose effect can be easily compensated by increase the number of $k'$.
These results also suggest that $R=100$ particles are sufficient to obtain an estimator with good precision and uncertainty.
While increasing $R$ can also decrease $k^{\prime}$ and thus reduce the total measurement times, but it is not recommended $k^{\prime}<6$ when $k=40$ because we find that a too small $k^{\prime}$ makes the estimation unstable, which is harmful to the precision.

\subsection{\label{sec:3-4}Uncertainty Bound}

Below we discuss the precision bound of our BPE-AL algorithm.
Instead of the Cramer-Rao lower bound often used in frequentist approaches, the precision bound of our algorithm can be given by the Ghosh bound~\cite{Ghosh1993}, which is proposed specifically for Bayesian approaches~\cite{Li2018}. 
The Ghosh bound for an estimated phase $\phi$ is defined as the reciprocal of the Fisher information~\cite{Holevo1982,Paris2009,Rubio2018} of a posterior probability distribution,
\begin{equation}\label{Ghosh_Bound}
  \Delta^2\phi_{\rm GB}(\vec{u}_{n}) = 
  \left(\int_{a}^{b}\mathrm{d}\phi\frac{1}{p(\phi|\vec{u}_{n})}
  \left(\frac{\mathrm{d}p(\phi|\vec{u}_{n})}{\mathrm{d}\phi}\right)^2\right)^{-1}.
\end{equation}
Thus, the Ghosh bound requires $\Delta^2\phi_{\mathrm{est}}(\vec{u}_n) \geqslant \Delta^2\phi_{\rm GB}(\vec{u}_{n})$, where the estimation uncertainty $\Delta^2\phi_{\mathrm{est}}(\vec{u}_n)$ is calculated from the variance of $p(\phi|\vec{u}_{n})$.

The calculation of the Ghosh bound can be implemented iteratively with $n$. 
Here, we apply central-limit theorem to approximate the posterior distribution to a Gaussian distribution to simplify the calculations in Eq.(\ref{Ghosh_Bound}). 
To be specific, a more strict binomial distribution likelihood function is used to describe $r$ of $R$ particles found in state $|1\rangle$ as~\cite{Dinani2019}:
\begin{equation}
  p(r|\phi ,\Phi) = \mathrm{C}_{R}^{r}
  p(1|\phi ,\Phi)^{r}
  p(0|\phi ,\Phi)^{R-r},
\end{equation}
where the two probabilities come from the likelihood function Eq.(\ref{Likelihood}).
This binomial distribution can be replaced by a Gaussian distribution when the total atom number $R$ is large enough~\cite{Dinani2019}:
\begin{equation}\label{center_limit}
  p(r|\phi ,\Phi)\approx
  \frac{1}{\sqrt{2\pi }\sigma }
  \mathrm{exp}\left\lbrack -\frac{{\left(r-R p(1|\phi,\Phi)\right)}^2}
  {2\sigma^2}\right\rbrack,
\end{equation}
where $\sigma^2 = R p(1|\phi ,\Phi)p(0|\phi ,\Phi)\approx r(R-r)/R$.
When $n>10$ the posterior probability can be approximated as a Gaussian function with negligible errors~\cite{Wiebe2016,Lumino2018,Danilin2018}.
Considering a posterior probability in the form of Eq.~(\ref{center_limit}), the reciprocal of posterior probabilities in Eq.(\ref{Ghosh_Bound}) can be simplified.
The result is shown in FIG.\ref{fig:FIG3}~(b) as blue points indicating that the uncertainty of the estimators with our BPE-AL obeys the Ghosh bound pretty well.

Finally, a simple comparison is made between our BPE-AL algorithm and a successful adaptive algorithm named the Gaussian optimization (GO)~\cite{Lumino2018}.
The update rule of $\Phi$ in the GO algorithm is designed to keep the variance of the summation of two likelihood functions Eq.(\ref{Likelihood}) taking its minimum in every update, resulting in optimal Bayesian updates and estimation efficiency.
The difference is that the update rule of $\Phi$ in GO algorithm is only design for phase estimation tasks where the binary data $u$ is provided by the single-particle system (scilicet the $R=1$ case).
For this reason in the comparison every result from GO are simulated by taking the average of $100$ single-paticle measurements.
After that the mean errors and stds of $50$ independent simulations are recorded, as well as the uncertainties calculated in the same way as above.
By contrast results from each iteration of our BPE-AL algorithm are provided in once where the particle number is set as $R=100$. 

The comparison results are shown in FIG.~\ref{fig:FIG3}~(c,d), where the x-axis is marked by the total measurement times $k^{\prime}N/k$.
The uncertainties of GO algorithm in (c) obeys standard quantum limit (SQL) while above the Ghosh bound evaluated with our BPE-AL algorithm.
In (d) for comparison of estimation precision, our BPE-AL algorithm reports mean values every $k^{\prime}=6$ points to keep the same consumption of particles.
With the same measurement times, our BPE-AL algorithm can outperform the GO algorithm.

\subsection{\label{sec:3-5}Robustness against Noises}

At last, we consider two kinds of noises often occurring in BPE experiments: depolarization noise and phase noise~\cite{Lumino2018,Rambhatla2020}.
Concretely speaking the depolarization noise caused by errors in measuring apparatus will result in omitted photon counts, 
and the phase noise caused by phase fluctuations will result in random errors in the adjustment of the auxiliary phase $\Phi$.
Here the depolarization and phase noises are simulated by adding Gaussian white noises on the origin values in Eq.(\ref{threshold_method}) and Eq.(\ref{Bayesian_update}), which are  respectively characterized by the noise strength parameters $q_d$ and $q_p$.
The values of readout population number $r$ and auxiliary phase $\Phi$ are then respectively changed to $r^{\prime}$ and $\Phi^{\prime}$ as:
\begin{equation}\label{noises}
  \begin{split}
    r^{\prime}&=(1+\kappa_d)r, \\
    \Phi^{\prime}&=\Phi + \kappa_p\pi, \\
  \end{split}
\end{equation}
where $\kappa_d$ and $\kappa_p$  respectively obey the Gaussian distributions as: 
$\kappa_d \sim \mathcal{N}(0, q_d^2)$ and $\kappa_p \sim \mathcal{N}(0, q_p^2)$.

\begin{figure}[t]
  \includegraphics[width = \columnwidth]{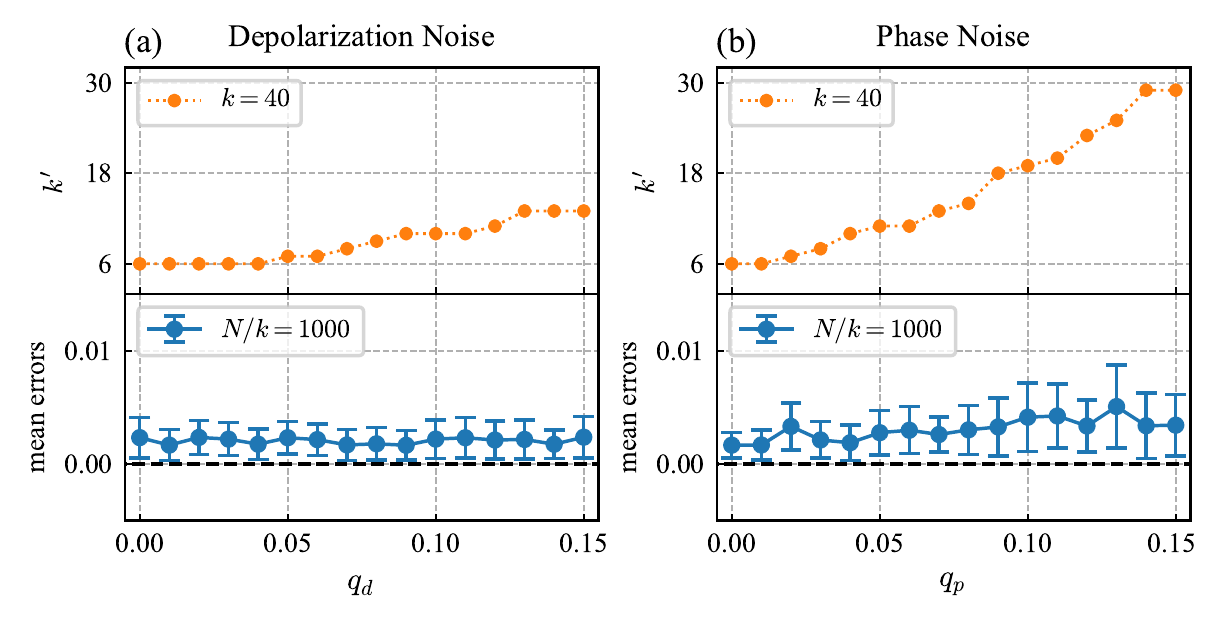}
  \caption{\label{fig:all_Noise}
  The performance in the presence of noises. The measurement times needed in each period $k^{\prime}$ versus the noise strength under (a) depolarization noises and (b) phase noises. Mean errors and stds with our BPE-AL algorithm under (c) depolarization noises and (d) phase noises. Here, $N/k=1000$ with $k=40$, and the estimated phase is $\phi=2.7624$.
  }
\end{figure}

Simulation results with our BPE-AL algorithm in the presence of noises are shown in FIG.~\ref{fig:all_Noise}.  
The measurement times needed in each period $k^{\prime}$ and the variations of mean errors and stds with noise strengths are given.
As the noise strengths $\kappa_d$ and $\kappa_p$ increase, the correspondingly measurement times used in each period $k^{\prime}$ have to be increased.
However, the mean errors and standard deviations do not change a lot when the strength of noises goes up.
This indicates that out BPE-AL algorithm has the ability to compensate the impact of noises by just increasing the measurement times.
Thus in the presence of noises, it is necessary to select more informative data to make a desirable estimation via our BPE-AL algorithm.

\section{Discussions}

We show how active learning can be used to implement a Bayesian phase estimation algorithm. 
Our algorithm involves an active learning process that can choose the relatively informative data and guide us to perform the Bayesian phase estimation with substantially less measurement times. 
For an example of two-level Ramsey interferometry, we adopt the entropy-based sampling method from active learning to find the significant auxiliary phases $\Phi$.
After the sampling process, for every $k$ Bayesian updates, only $k'$ (which is much smaller than $k$) measurements are needed to performed. 
Thus, in a process of large amounts of Bayesian updates, the real measurement times needed to perform in experiments can be reduced from $N$ to $k'N/k$ compared with the conventional BPE.
In the noiseless case, the typical measurement times with our algorithm can be $15\%$ of the one with conventional BPE. 
While in the presence of noises, by suitably increasing the measurement times, the performance of our algorithm can remain the same level. 
Excepting for reducing the measurement times, our algorithm also has the advantages of non-adaptive phase estimation algorithms such as perfect dynamic range and easy accessibility.

The entropy-based sampling method is used in the dataset generation part~\cite{Carleo2019} of BPE, meaning that reformative designs for BPE are also compatible to our algorithm.
For example, the Markov-chain Monte Carlo (MMC) or particle filter method~\cite{Granade_2012,Wang2017,Puebla2020} and particle guess heuristic (PGH) method~\cite{Wiebe2014} can be combined in, for the reason that we did no changes to the posterior update part of general BPE algorithms.
The particle filter method can be applied by simply shrinking the distance between the $k^{\prime}$ points in each period according to the narrowness of posterior functions, which on the other hand requires better phase adjusting precision.
The PGH method can be applied originally by adjusting the phase accumulation time to achieve better precision, as it is done in Ref.~\cite{Wiebe2016,Paesani2017}.

In realistic experiments, our algorithm is a promising alternative approach in real Bayesian parameter estimation tasks to simplify the experiment procedures such as atom clocks~\cite{Hume2007} and quantum magnetometers~\cite{Nusran2012}, by replacing the operations of realtime objective function maximization~\cite{Said2011,Waldherr2012,Bonato2016,Ruster2017,Dushenko2020,Dimario2020}, or providing an inherent adequate dynamic range without requirement of restarts setting in the middle of the estimations~\cite{Wiebe2016,Paesani2017}.
In addition, the active learning procedure in our algorithm is not relevant to the specific form of the likelihood function, suggesting that our algorithm can also work in situation where modeled effect of decoherence time $T_2$ are taken into account~\cite{Wiebe2016,Nusran2012,Paesani2017,Danilin2018}.
Furthermore, in varying parameters estimation cases the changing policy of $\Phi$ found by the active learning method can be optimized simultaneously through reinforcement learning.
Our algorithm provides a promising way for implementing efficient Bayesian phase estimation in various practical sensors. 

\begin{acknowledgments}
This work is supported by the National Natural Science Foundation of China (12025509, 11874434), the Key-Area Research and Development Program of GuangDong Province (2019B030330001), and the Science and Technology Program of Guangzhou (201904020024). M. Z. is partially supported by the National Natural Science Foundation of China (12047563). J. H. is partially supported by the Guangzhou Science and Technology Projects (202002030459).
\end{acknowledgments}

\section*{data availability}
  The data that support the findings of this study are available from the corresponding author upon reasonable request.

\bibliography{ActiveBayesian}
% Produces the bibliography via BibTeX.

\end{document}